\documentclass[letter]{aa} 
\usepackage{graphicx}
\usepackage{txfonts}
\usepackage{array}

\newcommand{\Ncr}{N_{\rm cre}}
\newcommand{\gme}{\gamma_{\rm e}}
\newcommand{\gmbr}{\gamma_{\rm br}}
\newcommand{\Ebr}{E_{\rm br}}

\newcommand{\cmc}{\rm cm^{-3}}
\newcommand{\ergcm}{\rm erg~cm^{-3}}
\newcommand{\dyncm}{\rm dyn~cm^{-2}}
\newcommand{\Egm}{E_{\gamma}}

\begin{document}

\title{Cosmic-ray electrons and the magnetic field of the North Polar Spur} 

\titlerunning{CRes and the magnetic field of the NPS}

\author{ Guobin Mou\inst{1},
         Jianhao Wu\inst{2,3},
         \and
         Yoshiaki Sofue\inst{4}
        }
       
\institute{ Department of Astronomy, School of Physics and Technology, Wuhan University, Wuhan 430072, China \\
      \email {gbmou@whu.edu.cn}
      \and
      Department of Physics and Institute of Theoretical Physics, The Chinese University of Hong Kong, Shatin, Hong Kong SAR, China \
      \email {jianhao.wu@link.cuhk.edu.hk}
      \and
      School of Computer Science, Wuhan University, Wuhan 430072, China \
      \and
      Institute of Astronomy, The University of Tokyo, Mitaka, Tokyo 181-0015, Japan
  }

\date{ }

\abstract
{} 
{The recent confirmation of the bipolarity of the eROSITA bubbles suggests that the well-known North Polar Spur (NPS)/Loop I probably is a 10 kpc sized relic in the Galactic halo and not a small local structure near the Sun. By virtue of multiwavelength observations of the NPS, unprecedentedly precise parameter constraints on the cosmic-ray electrons (CRes) and magnetic field in the post-shock halo medium can be provided. }
{ The parameters of the CRes and the magnetic field can be derived independently by modeling the gamma-ray and the radio data of the NPS via inverse Compton scattering and synchrotron emission, respectively. } 
{Our main results are (1) that the energy density of the CRe is (3--6)$\times 10^{-14}~\ergcm$, and the spectral index is $p\simeq 2.0\pm 0.1,  $ below the cooling break energy of about 5 GeV; 
(2) that the magnetic field strength is 3 $\mu$G; and
(3) that the shock acceleration efficiency of the CRe is (1--2)\%. 
Given the Mach number of 1.5, the high acceleration efficiency and flat spectrum of the CRe suggest that preexisting relativistic electrons may be reaccelerated in the NPS.  
Alternatively, these CRes could be accelerated by an evolving shock in the early epoch when its Mach number is high, and efficiently diffuse throughout the post-shock halo gas. In addition, the cooling break energy suggests that the cooling timescale is $10^7$ yr, which agrees with the age of the eROSITA bubbles. } 
{}
 

\maketitle

\section{Introduction}
Loop I, the giant loop spanning almost 100 degrees in the radio sky map, has been known for 60 years \citep{large1962}. Its eastern part is prominently brighter than its western part, and it is called the North Polar Spur (NPS). 
Loop I/NPS is outstanding in a wide range of frequencies from tens of MHz to tens of GHz (e.g., \citealt{deoliveira2008}), and the lower frequencies of $\lesssim 10^1$ GHz are thought to be dominated by synchrotron emission. 
Moreover, it is also visible in the X-ray  \citep{snowden1997} and gamma-ray band \citep{casandjian2009, su2010}.  
During the past decades, most works have regarded the Loop I/NPS as a local structure (LS) of $\sim 10^2$ parsecs that could originate from an old supernova remnant (SNR) or stellar activity \citep{weaver1979, wolleben2007}. 
A recent work \citep{panopoulou2021} investigated the optical polarization angles of nearby stars induced by foreground dust \citep{das2020}. The starlight polarization angles at Galactic latitude $b>30^{\circ}$ are essentially aligned with that of the radio NPS in tens of GHz, and based on this, Panopolou and collaborators argued that this part of the NPS should be located within $\sim 100$ pc. 

Nevertheless, Sofue thought that it could be a large halo structure (HS) of about 10 kpc at the Galactic center distance \citep{sofue1977, sofue2000, sofue2016} that might have originated from a past outburst of the Galactic center. This scenario is becoming attractive because it is consistent with the multiwavelength structures revealed in the recent decade, including their southern counterparts in the X-ray band \citep{predehl2020}, the Fermi bubbles \citep{su2010}, and the polarized radio lobes \citep{carretti2013}. 
In particular, the discovery of the southern eROSITA bubble provides compelling evidence that supports the HS picture, which is also supported by foreground absorption in the X-rays by the Aquila Rift clouds at a distance of 1 kpc \citep{sofue2015, sofue2016}. 
The NPS-like structure in the soft X-ray band is also reproduced in hydrodynamic simulations modeling the Fermi bubbles \citep{guo2012, mou2014, sarkar2019}. Because they overlap much, radio Loop I and the northern eROSITA bubble are probably the same physical structure, the radio emission of which probably come froms the synchrotron radiation of cosmic ray electrons (CRes) accelerated by the forward shock. 
The radio and X-ray NPS probably is a galactic center (GC)--distance halo structure, while it coincidentally overlaps with the foreground local dust and H I \citep{das2020}. As shown in our recent simulation study \citep{mou2023}, both the prominent east--west asymmetry of the Loop I/NPS and the faintness of its southern counterpart (north--south asymmetry), which are frequently quoted as support for the LS scenario, could be caused by a crossing CGM wind injected east by north in Galactic coordinates with a velocity of $\sim 200$ km s$^{-1}$ (see also \citealt{mou2018, sofue2019} for analytical studies).   
As inferred from X-rays, the Mach number of the Loop I/NPS is $\sim$1.5 (corresponding to a shock velocity of 300 km s$^{-1}$; \citealt{kataoka2013}), and the age of the Loop I/NPS probably is $10^7$ yr. In this context, the Loop I/NPS, which possibly has been misinterpreted for several decades, could be an excellent object for studying the physics of cosmic ray electrons and the magnetic field, and the particle acceleration of shocks and diffusion of CR on the galactic scale. 
The model and result are presented in Section \ref{S2}, and we discuss the results in Section \ref{S3}. 

\section{Model and result} \label{S2}
\subsection{Basic method}
For the radio NPS (see Figure \ref{fig1} for the 408 MHz map), our analysis is restricted to the latitude $b>30^{\circ}$ to avoid complications near the Galactic plane, which is also in line with the sky region for the gamma-ray data in \citet{johann2021}. 
The temperature spectral index $\beta$ ($T_B \sim \nu^{-\beta}$) of the NPS is $\sim$2.55 between 45 MHz and 408 MHz \citep{guzman2011}. It steepens as the frequency increases to GHz: $\beta \sim 2.8$ between 408 MHz and 2.3 GHz \citep{platania2003}, and $\sim 3.0$ between 408 MHz and 23 GHz \citep{miville2008}. \citet{vidal2015} found that $\beta$ of the NPS is about 3 between 23 and 41 GHz from WMAP data, but the dispersion is large, and \citet{jew2020} found it to be $ 3-3.2$ between 30 and 44 GHz from Planck data. The steepening of the synchrotron spectrum suggests a turning point at $\sim 10^0$ GHz, which is indicative of a cooling break in the CRe population. 
The main cooling mechanisms of CRes in the NPS involve synchrotron and inverse Compton scattering (ICS), which cause the spectral energy distribution (SED) of CRes to steepen by one power of $\gme$ for the continuous injection case.  
Here we assumed an SED of CRes to account for the NPS in a broken power-law form with the exponential cutoff at $\gamma_{\rm ct}$, 
\begin{equation}
~~~~~~~~~~~~ \frac{d\Ncr}{d\gme}= \left\{ 
\begin{aligned}
 N_0 \gme^{-p} ~~~~~(\gme < \gmbr)  ~~~~~~~~~~~~~~~~~~~~~~~ \\
 N_1 \gme^{-p-1} \exp(-\gme/\gamma_{\rm ct}) ~~~~( \gme \geq \gmbr) 
\end{aligned}
\right.
\label{ecr}
,\end{equation} 
where $\gmbr$ is the Lorentz factor of the cooling break energy $E_{\rm br}$. Continuity requires $N_1=N_0 \gmbr$. 
We investigated the cases with $p=1.9-2.2$, and set the exponential cutoff at $\gamma_{\rm ct}=9.8\times 10^5$ (0.5 TeV) for $p \leqslant 2.0$ and $1.96\times 10^6$ (1 TeV) for $p \geqslant 2.1$ to improve the fitting of the gamma-ray spectrum. 
For nonrelativistic bulk motion, which is the case here, the energy distribution peaks at $\gme \sim 2$ because the CRes follow a power-law distribution in momentum (instead of energy) with slope $p$ (\citealt{bell1978, sironi2013}).
Therefore, the SED of CRes (equation \ref{ecr}) can be regarded as starting from $\gme=2$ \footnote{When the lower bound of $\gme=1$ or 2 with $p \leqslant 2.2$ is adopted, the difference in the energy density of the CRes is only a few percent. }. The only two unknown parameters of the CRe population are $N_0$ and $\gmbr$ ($\Ebr$) in equation \ref{ecr}. 

We adopted one-zone assumption for simplicity. 
Because the energy density of the CRes and the magnetic field are coupled, these parameters cannot be derived from radio data alone. 
Thanks to the Fermi-LAT, the gamma-ray spectrum of the NPS was obtained (e.g., \citealt{johann2021}). The parameters of CRes can be solved independently via fitting the gamma-ray spectrum through ICS of CRe. After this, the magnetic field strength can be derived by fitting the radio spectrum. This is the basic method of this work. 

\subsection{Geometry structure}
\begin{figure}
\includegraphics[width=0.98\columnwidth]{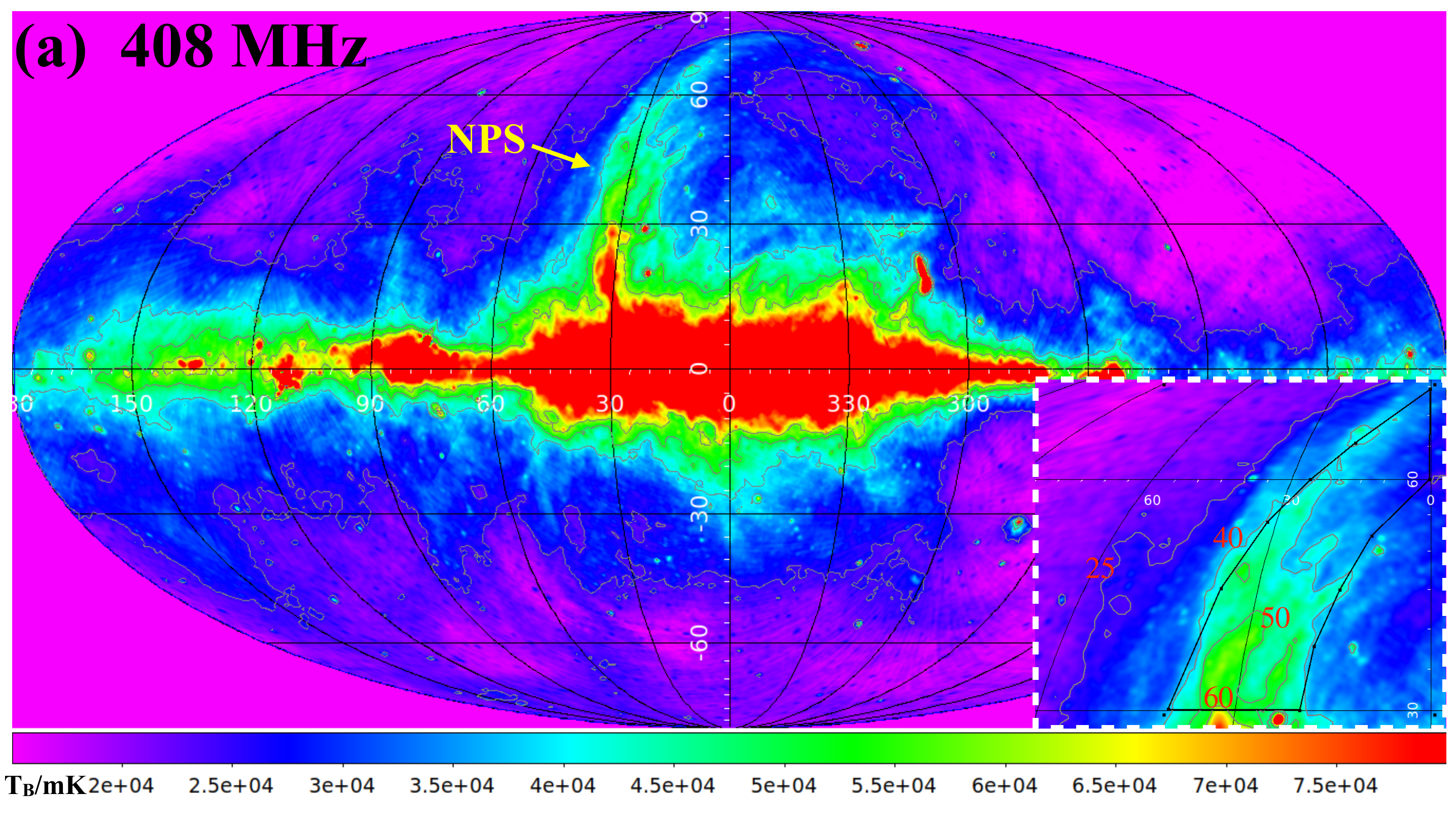}
\includegraphics[width=0.98\columnwidth]{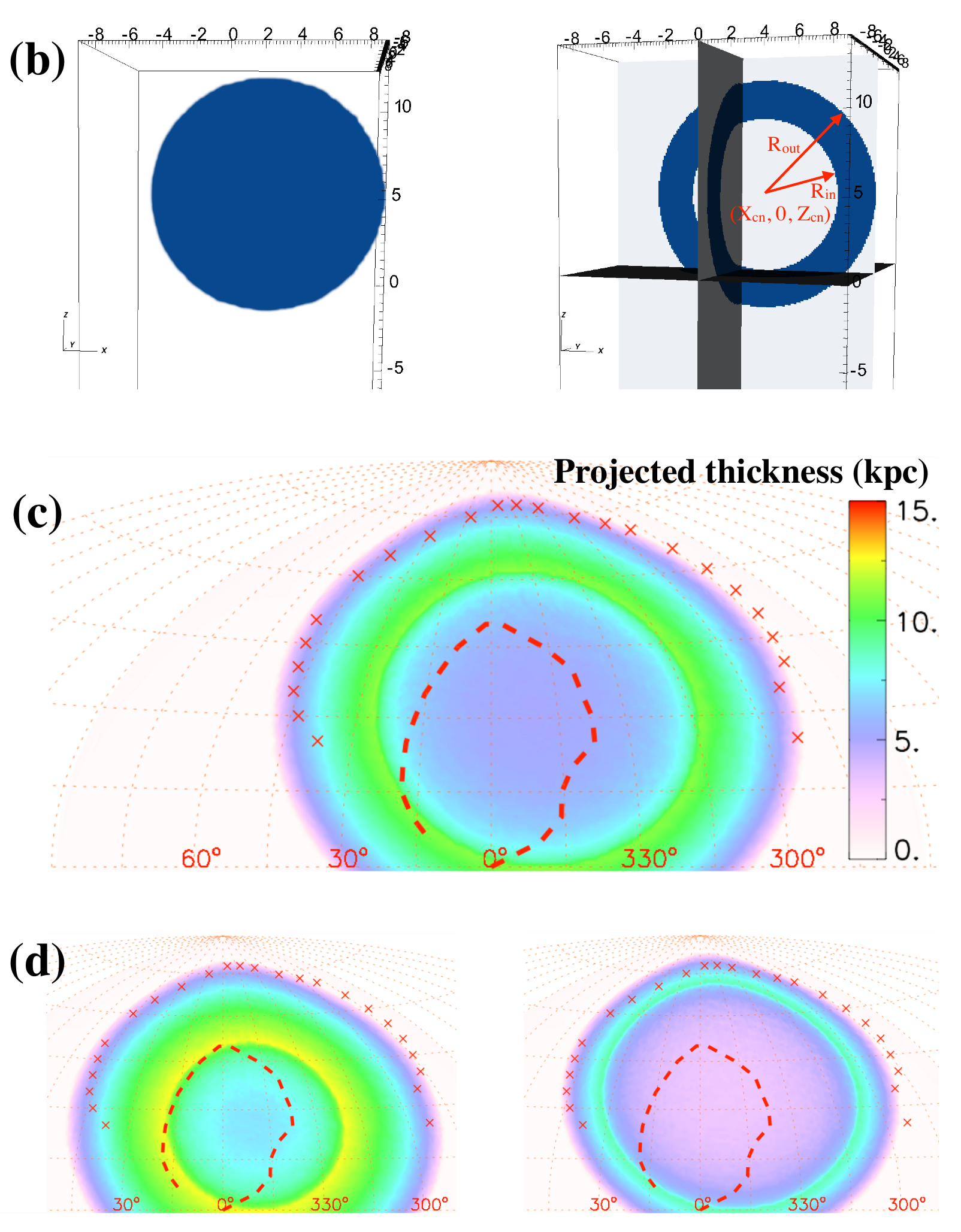}
 \caption{Observed NPS in the radio band and its geometry model. Panel (a) shows the 408 MHz sky map (\citealt{haslam1982} , $T_B$ in mK). The gray contours indicate $T_B=25$ K, 40 K, 50 K, and 60 K. The right window shows the zoom-in view of the NPS, and the polygon marks the region for radio intensity statistics (similar to the region for gamma-ray analysis in \citealt{johann2021}). Panel (b) shows the 3D view and the three-slice view of the shell (coordinate values are in units of kpc). Panel (c) shows the projected thickness ($D_s$) of the radiative shell, accounting for the Loop I/NPS in Galactic coordinates with $(R_{\rm in}$, $R_{\rm out})=$(5.0 kpc, 7.5 kpc). The crosses and the dashed line mark the outlines of Loop I and the northern Fermi bubble, respectively (see \citealt{su2010} for the coordinates). Panel (d) show the projected thickness maps for $(R_{\rm in}$, $R_{\rm out})=$(4.0 kpc, 7.5 kpc) (\emph{left}) and (6.0 kpc, 7.5 kpc) (\emph{right}). } 
 \label{fig1}
\end{figure}

The physical structure of Loop I/NPS is generally regarded as the post-shock medium. 
Before modeling the radio and gamma-ray NPS, we need to know the thickness of the radiative shell along the sightline -- $D_s(l,b)$, where $l/b$ is the Galactic longitude/latitude. 
We adopted 3D Cartesian coordinates, in which the $Z-$axis is the Galactic polar axis, and the Solar System is located at $(X,Y,Z)=(0, -8.2 ~{\rm kpc},0)$ \citep{bland2016}. 
We constructed a 3D hollow and thick-shelled bubble in the Galactic halo with its center at $(X_{\rm cnt}, 0, Z_{\rm cnt})$. The inner and outer radii of the shell are $R_{\rm in}$ and $R_{\rm out}$, respectively, in which we forced $R_{\rm in}=Z_{\rm cnt}$ to reduce the degrees of freedom. By testing a series of parameter groups, the four parameters can be estimated by comparing the shell projection in the Galactic coordinates and the observed NPS. 
The fitting result is $(R_{\rm in}$, $R_{\rm out})$=(5.0 kpc, 7.5 kpc) and $(X_{\rm cnt}, Z_{\rm cnt})=$(1.5 kpc, 5.0 kpc), and the modeled $D_s(l,b)$ is shown in Figure \ref{fig1}. The value of $D_s(l,b)$ depends on the specific direction, but is basically in the range of 6--10 kpc (Figure \ref{fig1}c). We adopted  $D_s(l,b)=8$ kpc as the fiducial value. 
For the LS scenario, when the center is 100 pc from the Sun, the projected thickness of the shell can be obtained by $D^{\rm LS}_s=D_s \cdot (100 {\rm pc}/9.4 {\rm kpc})=0.011 D_s=0.088$ kpc with the help of similar geometry. Thus, the values of the projected thickness under the two physical scenarios differ by two orders of magnitude, which has a significant impact on the parameters of the CRes that are required to fit the observations.  

\subsection{Gamma-ray NPS} \label{gamma}
There are two possible origins for gamma rays: ICS of the interstellar radiation field (ISRF) by CRes, or a hadronic origin ($pp$ collisions). If it is the hadronic origin, the gamma-ray luminosity per unit volume is 
$j_{\gamma} \sim f \sigma_{pp}e_{\rm crp} n_{\rm gas} c$, where $f\sim 0.17$ is the fraction of cosmic-ray protons (CRps) kinetic energy transferred into $\pi^0$ , which instantaneously decays into gamma rays, $\sigma_{pp}\sim 30$ mb is the cross section of inelastic collision \citep{aharonian2004}, and $e_{\rm crp}$ is the energy density of CRp. X-ray observations suggest that the density and temperature of the hot gas in the NPS are $n_{\rm gas} \sim (3-4)\times 10^{-3}~\cmc$ and 0.3 keV, respectively \citep{kataoka2013, kataoka2015}. If we had taken the thermal energy density of the ions as the upper limit of $e_{\rm crp}$, we would have a gamma-ray intensity of $J_{\gamma}=(4\pi)^{-1} j_{\gamma} D_s < 1\times 10^3$ eV cm$^{-2}$ s$^{-1}$ sr$^{-1}$, the upper limit of which is about half of the observed value \citep{johann2021}. Thus, a hadronic origin requires relatively extreme conditions in which the energy density of CRp could exceed that of thermal ions, and the gamma rays of the NPS more likely come from ICS. 

In ICS, the rate of gamma-ray production per unit volume per unit energy is given by 
 \begin{equation}
  \frac{dn_{\gamma}(\Egm)}{d\Egm}= c \int_{\gme} \int_{E_{ph}}  \frac{d\sigma_{\rm IC}(\Egm, E_e, E_{ph})}{d\Egm} \frac{d\Ncr}{d\gme} d\gme \frac{dn_{ph}}{dE_{ph}} dE_{ph}  
  \label{ics}
 ,\end{equation}
where $d\Ncr/d\gme$ is the energy distribution of the primary CRes (see equation \ref{ecr}), $dn_{ph}/dE_{ph}$ is the number density of an interstellar radiation field (ISRF) photon per unit energy. The differential cross sections of ICS in this equation can be approximated by 
$d\sigma_{\rm IC}/d\Egm= 3\sigma_{T} (E_e \Gamma_{\epsilon})^{-1} [2q \ln q+(1+2q)(1-q)+
(\Gamma_{\epsilon} q)^2 (1-q) (2+2\Gamma_{\epsilon}q)^{-1} ] $, where $\sigma_T$ is the Thomson cross section, $\Gamma_{\epsilon} \equiv 4 E_{ph} E_{e}/(m^2_e c^4)$, and $q \equiv \Egm \Gamma^{-1}_{\epsilon} (E_e-\Egm)^{-1} $ \citep{blumenthal1970}. 
The gamma-ray flux is found from the source term (equation \ref{ics}) as a sightline integral, 
\begin{equation}
\frac{dN_{\gamma}}{dE_{\gamma}}= \frac{1}{4\pi} \int \frac{dn_{\gamma}}{dE_{\gamma}} dR.
\end{equation} 

The ISRF takes the values in GALPROP \footnote{\url{http://galprop.stanford.edu}} v54 (see Figure \ref{figA1}, \citealt{moskalenko2006, porter2006}). The line of sight toward the NPS passes through different regions with different ISRFs. We took the ISRF at $(R, z)=$ (5 kpc, 5 kpc) as the seed photon ($R$ is the galactocentric distance, and $z$ is the height from the midplane of the Galactic disk). 

\subsection{Radio NPS}
For the radio NPS  (Figure \ref{fig1}a), we selected a similar region to the one used in the gamma-ray analysis \citep{johann2021}, and simply chose the region of longitude $l=60^{\circ}$ (outside and close to the NPS) as the radio background or foreground, of which $T_B \sim 22$ K at 408 MHz. 
Subtracting the background or foreground of the same latitudes, we derived that the average brightness temperature of the NPS at 408 MHz is $T_B \simeq 20$ K. 
By fitting the gamma-ray spectrum, we obtained a series of possible distributions of the CRe with different $\Ebr$ and $N_0$. For each SED, we then calculated the synchrotron emission with different magnetic field strengths, in which the pitch angle between the electron velocity and the field was assumed to be random. 
We defined the radio-fit conditions as follows: matching the intensity at 408 MHz, spectral indices of $\bar{\alpha}_1=-0.55\sim -0.60$ (45--408 MHz), and $\bar{\alpha}_2=-1.0\pm 0.1$ (0.408--23 GHz), in which $\bar{\alpha}_1 \equiv \ln (S_{45 {\rm MHz}}/S_{408 {\rm MHz}})/\ln (45/408)$ and $\bar{\alpha}_2 \equiv \ln (S_{408 {\rm MHz}}/S_{23 {\rm GHz}})/\ln (0.408/23)$. 
These conditions restrict the model parameters to a narrow range.  

\begin{figure*}
\centering
\includegraphics[width=1.90\columnwidth]{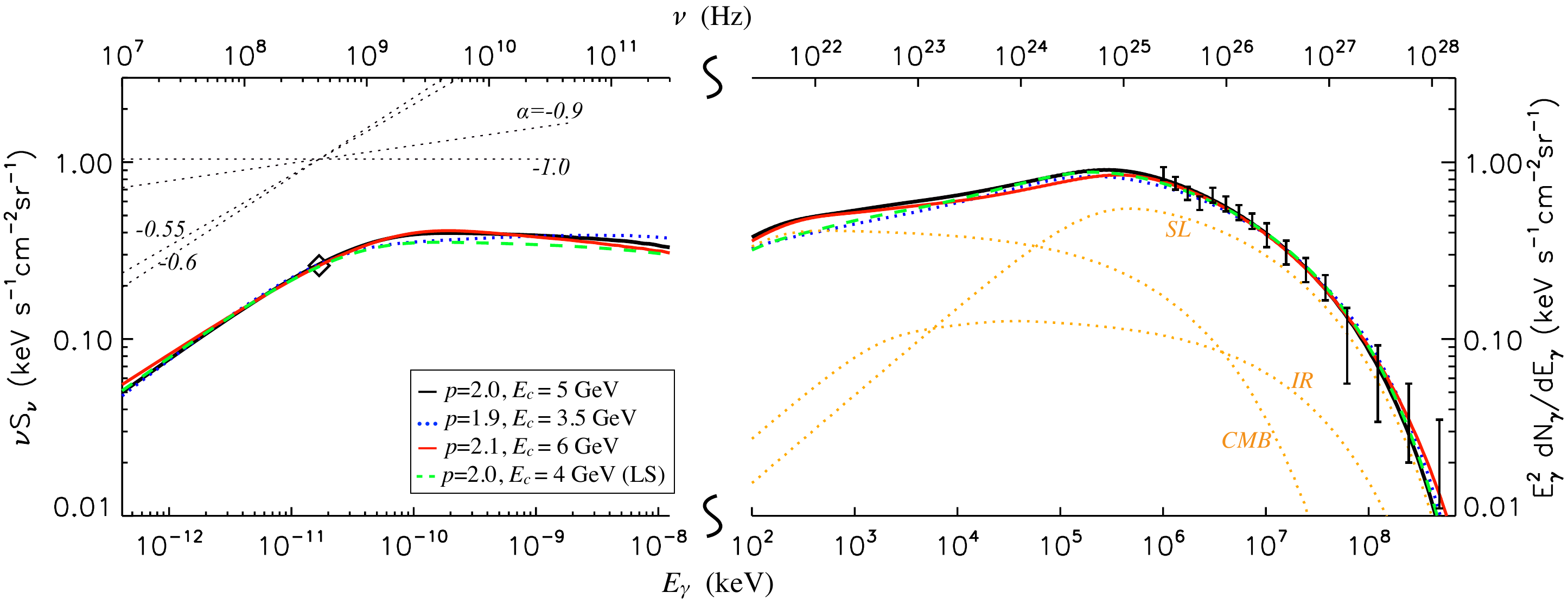} 
 \caption{ Modeling the radio and gamma-ray spectra for the NPS. 
The diamond and error bars indicate the radio intensity (408 MHz) and gamma-ray data \citep{johann2021} of the NPS, respectively. The dotted orane lines indicate the contributions of different components of the ISRF for $p=2.0$ (see figure \ref{figA1} for details).  } 
 \label{fig2}
\end{figure*} 

\begin{table}
\centering
\caption{Parameters for modeling the radio and gamma-ray NPS. A Y or N (yes or no) after the $\Ebr-$value indicates whether the values of the modeled $\bar{\alpha}_1$ and $\bar{\alpha}_2$ meet the radio-fit conditions, and LS after the $p-$value marks the local structure scenario. } 
\setlength{\tabcolsep}{0.14cm} {
\begin{tabular}{ccccccc}
\hline \hline
$p$ & $\Ebr$ & $N_0$    & $e_{\rm cre}$   & $B$  & $\bar{\alpha}_1$ & $\bar{\alpha}_2$  \\
   & GeV   & $\cmc$ &  $\ergcm$  &  $\mu$G  &   &    \\
\hline
1.9     & 3.5 (Y) & $2.7\times 10^{-9}$ & $3.4\times 10^{-14}$ & 3.4 & -0.57 & -0.91 \\ 
2.0    & 4 (N)   & $7.8\times 10^{-9}$ & $5.6\times 10^{-14}$ & 2.9 & -0.61 & -0.95 \\ 
2.0    & 5 (Y)   & $6.3\times 10^{-9}$ & $4.6\times 10^{-14}$ & 3.2 & -0.57 & -0.91 \\
2.0    & 6 (N)   & $5.2\times 10^{-9}$ & $3.9\times 10^{-14}$ & 3.4 & -0.55 & -0.87 \\
2.1     & 6 (Y)   & $1.4\times 10^{-8}$ & $6.0\times 10^{-14}$ & 3.3 &  -0.60 & -0.91 \\ 
2.2    & 8 (N)   & $3.4\times 10^{-8}$ & $9.5\times 10^{-14}$ & 3.2 &  -0.63 & -0.89 \\
1.9LS  & 3 (Y)  & $1.8\times 10^{-7}$ & $2.3\times 10^{-12}$ & 4.1 & -0.59 & -0.91 \\
2.0LS & 4 (Y)  & $4.7\times 10^{-7}$ & $3.4\times 10^{-12}$ & 3.7 & -0.59 & -0.93 \\
2.1LS  & 5 (Y) & $9.9\times 10^{-7}$ & $4.3\times 10^{-12}$ & 3.9 & -0.60 & -0.93 \\
2.2LS & 8 (N) & $2.1\times 10^{-6}$ &  $5.8\times 10^{-12}$ & 4.0 & -0.62 & -0.87 \\
\hline \hline  
\end{tabular} }  
\label{tab:tabl1}
\end{table} 

\subsection{Results}
We modeled the radio and gamma-ray data with the setup with $p=2.0$ and $\Ebr=$5 GeV as the fiducial case (Table 1 and Figure \ref{fig2}). The coefficients of the CRe SED (equation \ref{ecr}) in the fiducial case were $N_0=6.3\times 10^{-9} ~\cmc$ and $N_1=6.2\times 10^{-5} ~\cmc$. The energy density of the CRes is thus $e_{\rm cre}=4.6\times 10^{-14} ~\ergcm$, and the number density of the CRes ($\gme \geqslant 2$) is 
$3.3\times 10^{-9}~\cmc$, which is $10^{-6}$ of the thermal density derived from X-ray observations \citep{kataoka2013}. In addition, the field strength is around 3 $\mu G$, and the energy density of the CRes is about 11\% of that of the magnetic field, indicating a significant deviation from energy equipartition between the CRes and the magnetic field. 

We also derived the confidence intervals of the CRe parameters beyond which the radio-fit conditions will never be met: $\Ebr \simeq$ 3.5 GeV for $p=1.9$, 4--6 GeV for $p=2.0$, and 6 GeV for $p=2.1$ (Table 1). The radio conditions rule out the cases of $p \geqslant 2.2$ and $p \leqslant 1.8$. Thus, the possible range of $e_{\rm cre}$ would be $(3.4-6.0)\times 10^{-14}$ erg cm$^{-3}$. 

The gamma-rays of the NPS are dominated by ICS on the starlight (SL; Figure \ref{fig2}). This is different from the Fermi bubbles, where most of the ICS signals below $\sim 100$ GeV come from the cosmic microwave background (CMB) in the leptonic model \citep{ackermann2014} because the energy of CRes needs to exceed 300 GeV to generate GeV photons by scattering the CMB photons, and the these electrons are quite rare in the NPS because the SED steepens above the cooling break energy $\Ebr$. 
Since the starlight dilutes with the height from the Galactic disk, this agrees with dimming of the gamma-ray NPS with latitude (e.g., \citealt{ackermann2014}). 

\section{Discussion} \label{S3}
According to X-ray observations, the post-shock gas accounting for the X-ray NPS has a temperature of 0.3 keV and a density of $n_{\rm gas} \simeq (3-4)\times 10^{-3}~{\rm cm^{-3}}$ \citep{kataoka2015}. Thus, the thermal pressure is $P_{\rm th} \simeq 3 \times 10^{-12}~\dyncm$. The CRe acceleration efficiency of the shock 
in our fiducial case is 
$\eta_{e} \simeq e_{\rm cre} v_2/\left[\frac{1}{2}\rho_1 v^3_s (1-\mathcal{C}^{-2}) \right]=$1.8\%, where $\mathcal{C}$ is the compression ratio, $v_2$ is downstream velocity, $v_s$ is the shock velocity, and $\rho_1$ is the upstream density. This suggests that CRes can be efficiently accelerated by weak shocks. When the confidence intervals are considered, $\eta_e$ would be (1--2)\%.  

This value is unusually high compared with the expectation of the canonical diffusive shock acceleration (DSA) theory
given the low Mach number (e.g., $\lesssim 10^{-4}$ for Mach number $\lesssim 3$; \citealt{kang2013}), but agrees with the speculation from observations of some radio relics. 
For these radio relics, the observed radio brightness suggests that $\eta_e$ is probably very high \citep{kang2012, brunetti2014, vazza2015, botteon2020}, which cannot be reconciled with the DSA mechanism (but see also \citealt{locatelli2020} for a radio relic consistent with DSA).  

Moreover, the Mach number of 1.5 inferred from X-rays \citep{kataoka2013} is significantly lower than that deduced from DSA for $p < 2.2$ ($\mathcal{M} > 4.6$).
Nevertheless, a similar case also arises in some radio relics where the X-ray derived Mach numbers are significantly lower than those inferred from radio spectra (see reviews by \citealt{brunetti2014, vanWeeren2019}).  

For the low Mach number, one could argue that, the electron temperature in the post-shock halo gas from X-rays might be lower than the ion temperature, and the Mach number might therefore be underestimated. 
Because of the high ion-to-electron mass ratio, the post-shock ions (protons) are initially heated to $T_{p,0}\simeq 3 m_p v^2_s/(4k_B \mathcal{C}),$ 
where $v_s$ is the shock speed, while electrons are heated to $T_{e,0}\simeq T_{p,0} (m_e/m_p)$. Then, electrons gain energy while ions lose energy in Coulomb collisions. In the early period of this process, the ion temperature does not change significantly, and the evolution of the electron temperature \citep{spitzer1962} can be simplified to 
\begin{equation}
T_e(t)\simeq \max \left[3.6\times 10^5 ~{\rm K} \left(\frac{n_p}{1~{\rm cm}^{-3}} \frac{t}{1~\rm yr} \frac{T_{p,0}}{10^7 K}\right)^{2/5}, T_{e,0}\right] ~.  
\end{equation}
The equipartition timescale between electrons and ions is $t_{\rm eq} \simeq 0.72~ {\rm Myr} \left(\frac{T_{p,0}}{10^7~K} \right)^{3/2} \left(\frac{n_p}{10^{-3}{\rm cm}^{-3}} \right)^{-1}$ \footnote{Accurate calculations show that the required time is $3.6 t_{\rm eq}$ for the temperature difference between electrons and ions to fall within 10\%. }, while the dynamic timescale is $t_{\rm dyn}\equiv R_{\rm NPS}/v_s \gtrsim 15{\rm Myr} \left(\frac{R_{\rm NPS}}{10~{\rm kpc}}\right) \left(\frac{T_{p,0}}{10^7~K}\right)^{-1/2}$, where $R_{\rm NPS}$ is the current size of the NPS. 
For the NPS with the X-ray-inferred temperature of 0.3 keV, we find that Coulomb collisions alone can efficiently heat electrons up to the temperature of ions, and thus the Mach number derived based on the thermal equilibrium is self-consistent.   

We speculate that the high acceleration efficiency of CRe and flat radio spectra suggest that the CRes might be reaccelerated in the NPS. 
The CRes could have experienced multiple rounds of reacceleration by multiple weak shocks, and each reacceleration flattens the spectrum \citep{melrose1993,kang2021}. Observational signs of multiple bursts in the Galactic center over the past million years have been reported \citep{bland2013,bordoloi2017}. 

Nevertheless, there is another scenario according to which, the majority of CRe may have been accelerated by an evolving shock with a higher Mach number (e.g., $\mathcal{M}\geq 5$ ) in the early epoch, and transported to the current position, whereby the physics of the CRes is not related with the current Mach number. 
Given the enclosed mass of the halo medium ($M(r)\simeq 1\times10^7 M_{\odot} r^{1.5}_{\rm kpc}$; \citealt{miller2015}), the radius of the evolving shock when $\mathcal{M}\geq 5$ is $\lesssim 0.2R_{\rm NPS}\simeq2$ kpc via the approximate relation that the shock energy $\sim M(R_s) ~\dot{R}^2_s \sim$ constant. 
From the conservation of mass, we derives that at present, the gas swept by a shock with $\mathcal{M}\geq 5$ constructs a 0.3 kpc shell adjacent to the contact discontinuity that is separated from the current shock front by the gas swept by that shock with a Mach number falling below 5 ($5>\mathcal{M}\geq 1.5$). The CRes within must diffuse through a thickness of several kiloparsecs to the current shock front. 
When we consider that the projected direction of the magnetic field basically traces the pattern of the NPS \citep{planck2016}, the perpendicular diffusion coefficient in the evolving-shock picture probably is comparable to or exceeds the isotropic coefficient by several times $10^{28}~{\rm cm^{2} ~ s^{-1}}$ \citep{strong2007}.  
Further discussion of this picture is beyond the current one-zone approximation in this Letter and is left for future work.  

The cooling break energy of CRes ($\Ebr$) deduced from the radio and gamma-ray spectra is about 5 GeV. Given the ISRF energy density of $1.27\times 10^{-12}~ \ergcm$ at $(R,z)=$ (5 kpc, 5 kpc) and the field strength of 3 $\mu$G, the cooling timescale for CRes of 5 GeV is 60 Myr. 
When we adopt an ISRF energy density of $2.3\times 10^{-12}~\ergcm$ near the midway of $(R,z)=$ (3 kpc, 3 kpc), the cooling timescale would be shortened to 30--40 Myr. Thus, the cooling timescale of the cooling break energy agrees with the eROSITA bubble ages of 20 Myr \citep{predehl2020}, suggesting that the results agree with the premise of the halo-structure nature. 

For the LS scenario, the energy density of CRe we deduced is $(2-4)\times 10^{-12} ~\dyncm$.  
As a comparison, the pressure of the hot plasma of $10^6$ Kelvin filling the local hot bubble is estimated to be $\sim 1\times 10^{-12}$ dyn cm$^{-2}$ \citep{puspitarini2014, snowden2014}.  
The magnetic field measured by Voyager 1 when it crossed the heliopause is $\sim 5 ~\mu$G \citep{burlaga2014}, while the estimate based on modeling of the radio emission in the Galactic plane is that the magnetic field at the position of the Sun is $\sim 3 ~\mu$G \citep{jaffe2010}. When we take the magnetic field of 5 $\mu$G as the representative value of the local bubble and assume that CRs and the magnetic field are in energy equipartition, the total pressure (including thermal pressure, magnetic pressure, and CRe pressure) of the local bubble is $\sim 3 \times 10^{-12}$ dyn cm$^{-2}$, which is close to the value in \citet{cox2005}. 
Thus, the acceleration efficiency of the CRes in the LS scenario would be unusually high, which is a challenge for our current knowledge of the shock acceleration of SNRs. 
In addition, by integrating over the volume in the LS scenario (see the shell in Figure \ref{fig1}b, but scaled down 0.011 times), the total CRe energy exceeds $1\times 10^{50}$ erg. This value is also extraordinary compared with the $10^4$ yr old SNRs Cygnus Loop and W44, whose total CRe energies derived from modeling the radio and gamma-ray data are $< 1\times 10^{49}$ \citep{katagiri2011} and $\sim 10^{48}$ erg \citep{ackermann2013}, respectively.  


\begin{acknowledgements}\small
We thank the anonymous referee for insightful comments that improved our work. GM thanks Dr. Zhongqun Cheng and prof. Ruizhi Yang for helpful discussions. G.M. is supported by the National Program on Key Research and Development Project (Grants No. 2021YFA0718500, 2021YFA0718503), and NSFC (nos. 12133007 and 11833007).   
\end{acknowledgements}


\appendix
\setcounter{figure}{0} 
\renewcommand{\thefigure}{A\arabic{figure}}
\setcounter{table}{0}
\renewcommand{\thetable}{A\arabic{table}}

\section{Additional figure}

\begin{figure}[h]
\includegraphics[width=0.97\columnwidth]{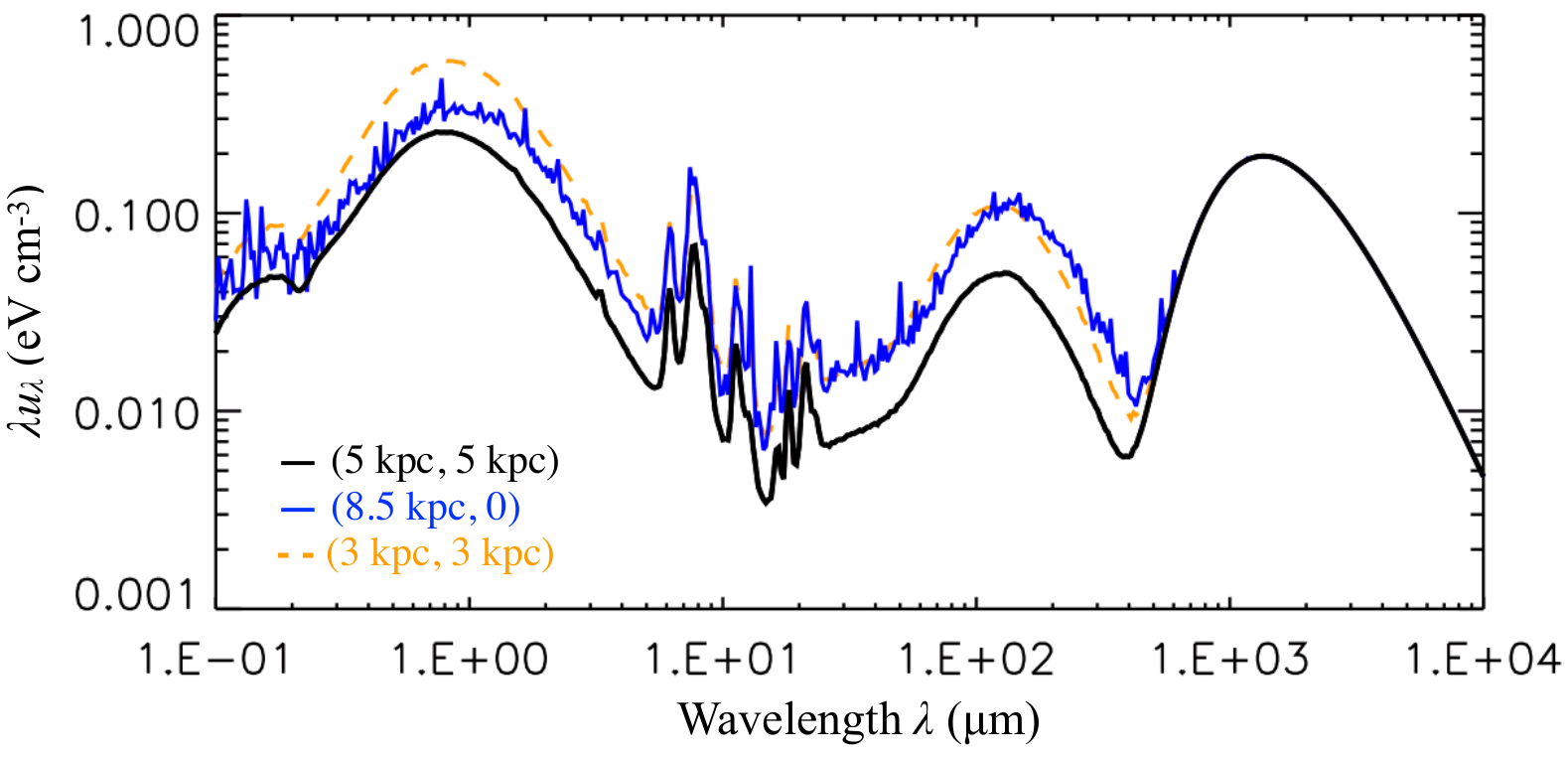} 
 \caption{ ISRF at three different positions: $(R, z)$=(5 kpc, 5 kpc), (8.5 kpc, 0), and (3 kpc, 3 kpc). The three different components in the spectra are clearly visible: starlight, infrared light, and the CMB. } 
 \label{figA1}
\end{figure} 


\end{document}